\documentstyle[aps,prl,multicol,epsf]{revtex}

\begin{document}

\title{Dynamics of directed graphs: the world-wide Web}

\author{Bosiljka Tadi\'c}

\address{Jo\v{z}ef Stefan Institute,
P.O. Box 3000, 1001 Ljubljana, Slovenia }


\maketitle
\begin{abstract}
We introduce and simulate  a growth model of the world-wide Web based on
the dynamics of outgoing links that is motivated by the conduct of the
agents in the real Web to update outgoing links (re)directing them towards
constantly changing selected nodes.   Emergent statistical correlation
between the distributions of outgoing and incoming links is a key feature
of the dynamics of the Web. The growth phase is characterized by temporal
fractal structures which are manifested in the hierarchical organization
of links. We obtain quantitative agreement with the recent
empirical data in the real Web for the distributions of in- and out-links
and for the size of connected component.  In a fully grown network of $N$
nodes we study the structure of connected clusters of nodes that are
accessible along outgoing links from a randomly selected node.
The distributions of size and depth of the connected clusters with a
giant component exhibit supercritical behavior. By decreasing the control
parameter---average fraction  $\beta $ of updated and  added links per
time step---towards $\beta _c(N) < 10\% $ the Web can resume a critical
structure with no giant component in it. We find a different universality
class  when the updates of links are not allowed, i.e., for $\beta \equiv 0$,
 corresponding to the network of science citations.
 \end{abstract}

\begin{multicols}{2}
\section{Introduction}
Understanding the evolution of complex networks
\cite{cell,socio,sci,internet,HTTP,earlier_exp2,surfing,access,classes,BJA,DMS,other,RandomGraph}
 remains  a challenging problem of
modern statistical physics. Modeling  the dynamic  structure of these
networks has significant practical applications, for instance in the
design of  search and control algorithms and in predicting emergence of
new phenomena in  information networks \cite{HTTP}.
An evolving  network consists of increasing number of nodes, which are
connected by links
according to certain rules specific to each network.  Although the spatial
distribution of nodes in a network is random, a different structure
can be observed when the distribution of links is considered.
In this respect two major groups \cite{classes} are:
(1) random networks, with the number of links attached to a node
fluctuating around an average value, and (2) networks with  hierarchical
structure of links, manifested in the power-law decay of the distributions
of node ranks.
The structure of connections  has immediate impact on the accessibility
of nodes and on functional stability of the network. For instance,
the hierarchically connected networks are shown \cite{BJA} to be robust
to large failure rates, however, they are vulnerable to removal of a
few 'key nodes'.
The inhomogeneous structure of connections has been  found  in
metabolic cycles \cite{cell}, in various  social networks \cite{socio},
including the science citation network \cite{sci}, and the Internet
\cite{internet} and the world-wide Web
\cite{HTTP,earlier_exp2,surfing,access}.

The world-wide Web is not a separate physical networks. Rather it is
a subset of the Internet that uses a specific protocol (hypertext transfer
protocol) and directed links to access a variety of data,
representing an example of distributed client/server computing \cite{java}.
Based on these properties a hierarchical structure of links emerges that
is peculiar to the world-wide Web.
Recently large-scale  measurements in  the world-wide Web
that  involved  $10^8$ nodes  reveal \cite{HTTP} that: (a)
Both the distributions of incoming links and distribution of
outgoing links follow a power-law behavior with  different
exponents;
(b) The distribution of size of connected component---number of
linked nodes in the web crawls---also exhibits scaling behavior; and (c)
The Web shows a very intricate structure of connections, with a large
strongly connected component in the center.

Recently a great deal of effort  was devoted to understand
emergence of the  scale-free structure of links in various complex
networks \cite{BJA,DMS,other,RandomGraph}.
An important ingredient of the dynamics---preferential attachment \cite{BJA}
 was shown to lead to the power-law distribution for incoming links
(the outgoing links  are  fixed by the rules of the model \cite{BJA}).
The model of preferential attachments
for incoming links has been generalized to include initial
attractiveness of nodes  in Ref.\ \cite{DMS},
where the authors  proved  analytically the existence of the scaling
region  for large evolution times. Recently it was
demonstrated in Ref.\ \cite{RandomGraph} how a giant connected component
can occur in the model  of random directed graphs assuming arbitrary and
statistically independent distributions of incoming and outgoing links.
The predictions of the random graph theory, however, show
substantial quantitative differences  compared to the
properties of the real world-wide Web \cite{RandomGraph}.

In this work we simulate growth and response of a directed graph
with the dynamic rules that are motivated by prominent features of the
world-wide Web.  These dynamic rules  yield the statistically correlated
distributions of outgoing and incoming links. Another
important feature of the model is that
 the links between pairs of nodes are not fixed in time, but may
vary on the time scale of the network's evolution (updates of links).
This property, that  may be found in some metabolic cycles, is
not shared with many other social networks,
whose physical links are  either fixed or vary on much slower time scale.
We argue that the (at present high) frequency of  updates of the outgoing
links, which is peculiar to the agents in the Web, is essential for the
observed inhomogeneous structure of connections. We find a consistent
agreement with the recent empirical data on the real  world-wide Web both
in the structure of links and in the response of the network.

The  paper is organized as follows: In Sec.\ II we introduce the
growth rules of the network and determine the emergent rank distributions
of outgoing and incoming links. In Sec.\ III we study the
temporal fractal structures in the growth phase of the
network and determine the corresponding distributions of the first-return
links. Sec.\ IV is devoted to investigation of the connected components on
the fully grown network and fractal properties of noise. A short summary and
the discussion of the results is given in Sec.\ V.


\section{Growth model  and rank distributions}

We suggest the  following  simplified set of rules to incorporate
basic features pertinent to the real world-wide Web:
{\bf (i) Directed nature of linking.}
The world-wide Web represents  a {\it directed graph} with nodes
corresponding to Web pages, and arcs corresponding
to hyperlinks between pages \cite{HTTP,RandomGraph}.
 {\bf (j) Growth and rearrangements at unique time scale.}
At each time unit $t$ a new node $i=t$ is added to the network
 ({\it growth}) and a number $M(t)$ of new links are distributed
among the nodes ({\it update of links}) following two rules specified below.
A fraction $f_0(t)\equiv \alpha M(t)$ of new links
are outgoing links from the new added node $i=t$, whereas the rest
$f_1(t)\equiv (1-\alpha )M(t)$ are the updated links at other nodes
in the network. Hence, the relevant parameter in the model
 is the ratio of updated and added links at
each time step, i.e., $\beta \equiv f_1(t)/f_0(t) =(1-\alpha )/\alpha$,
which is independent of the actual number of links $M(t)$.
Thus, in the model $M(t)$ represents a net increment of the number of
outgoing links at time step $t$. Variations in $M(t)$ are, in principle,
not restricted, being caused by adding  new connections between
old nodes (resulting in the increase of the number of links), and/or by
 removing some earlier links (causing decrease of the number of links).
We assume that  variations in $M(t)$ are such that we can define
 an average value $M\equiv {\overline{M(t)}}$,
which can be considered as a constant in first approximation. In practice,
the number of nodes {\it and} the number of links in the network increases
with time, so that reasonable values for the average $M$ are positive.
For consistency, we keep $M=1$ throughout this work.
Different values of $M$ do not affect the universal properties in the scaling
region (i.e., for large evolution times). Taking much larger $M$
values, however, considerably increases computation time.
{\bf (k) Preferential update (k1)  and preferential attachment (k2).} These
properties represent a paradigm of social behavior, for instance preferential
attachments  are driven by ``popularity'' of a node, as discussed in earlier
models of evolving social networks. In addition, here we have that
not all  nodes are getting updated at every time step, rather
only a few nodes update at a time. Moreover, some of the nodes update
outgoing links more frequently than others. This features can be formulated
in terms of probabilities as follows:
Where do the links come from?
We assume that apart from the new node, updating at time $t$ occurs with
larger probability at most active nodes ({\it preferential
activity}), i.e., an outgoing link from
the node $k\le i$ appears with the probability
 \begin{equation}
 Prob1 = {{\alpha M + q_{out}(k,t)}\over{(1+\alpha)M*i}} \ .
\label{outlinking}
\end{equation}
A new link goes to the node $n$ (whose age is $t-n$) with
the probability
\begin{equation}
 Prob2 = {{\alpha M + q_{in}(n,t)}\over{(1+\alpha)M*i}} \ ,
\label{inlinking}
\end{equation}
i.e.,  a ``popular'' node attracts even more links.
 Here $q_{out}(n,t)$ and $q_{in}(n,t)$ are the dynamically varying
number of outgoing and incoming links, respectively, at the node
$n$ at current time $t$. It is assumed that at the time of addition of
a node $i$ to the network $q_{out}(i,i)=q_{in}(i,i)=0$.
Note that the second rule in Eq.\ (\ref{inlinking}) is formally
equivalent to the  model of
preferential attachment for {\it incoming} links \cite{DMS}.
However, in our model the role of the parameter $\alpha $ in this equation
is precisely determined through the rule in Eq.\ (\ref{outlinking}),
which regulates {\it outgoing} links:
(1) For $\alpha \equiv 1/(1+\beta )< 1$  a
fraction $(1-\alpha )M$ of outgoing links in Eq.\ (\ref{outlinking})
refers to updates at earlier nodes;
(2) For $\alpha =1$ (equivalent to $\beta =0$) no updates are allowed.
 New links originate only from the new added node;
(3) Formally one can apply Eqs.\ (\ref{outlinking}-\ref{inlinking})
for $\alpha $ strictly larger than unity,
however, in this region $\beta $ becomes negative, suggesting
 that the number of links in the network is not conserved.
Therefore only the incoming links (\ref{inlinking}) can be counted
correctly when $\alpha $ is in the interval $\infty > \alpha >1$;
(4) In the mathematical limit $\alpha \to \infty $, or in practical
cases when $\alpha \gg {\bar{q}}/M$, the  effects of the dynamical
quantities $q_{out}(k,t)$ and $q_{in}(n,t)$ become negligible,
and we recover the case of fully random network.

In what follows we would like to demonstrate that the features (i), (j),
(k1) and (k2) explained above are essential for the behavior observed
in the Web.
Earlier models \cite{BJA,DMS} concentrated only on the preferential
attachment rule (k2), while largely neglecting properties (i), (j) and (k1).
The most complete account of the preferential attachment
model is  given in Ref.\ \cite{DMS}. On the other side, the random graph
model \cite{RandomGraph} considered the directed nature of network (i),
while neglecting the details of the linking rules (j), (k1) and (k2).

Using  the  rules in Eqs.\ (\ref{outlinking})-(\ref{inlinking})
we grow a large network of $N=10^6$ nodes.  We measure the ranking of
nodes in this network by the distributions of outgoing links and
incoming links,  $P(q_{out})$ and  $P(q_{in})$, respectively, which are
shown in Fig.\ 1 for
parameter $\beta = 3$, corresponding to $\alpha =0.25$ in Eqs.\
(\ref{outlinking}-\ref{inlinking}).
For comparison we also simulate the distributions of outgoing and incoming
links in the case of fully random directed network.
For a range of values of their arguments the {\it cumulative} distributions
for finite $\beta $ values exhibit a power-law behavior according to
\begin{equation}
P(q_{out}) \sim q_{out}^{-(\tau _{out} -1)} \ ;\ \
P(q_{in}) \sim q_{in}^{-(\tau _{in}-1)} \ .
\label{PP}
\end{equation}
By increasing the parameter $\alpha $ in Eqs.\
(\ref{outlinking})-(\ref{inlinking}) in the range $(0,1)$, corresponding
to decrease of the relative fraction of updated links $\beta $ in
the interval $(\infty,0)$, the slopes of the distributions increase.
The scaling exponents $\tau _{in}$ and $\tau _{out}$ given in
the inset to Fig.\ 1  are  parametrized by $\alpha $. Our numerical
results  for  $\tau _{in}$  agree with the
exact solution \cite{DMS}  $\tau _{in} = 2 +\alpha $  for the incoming links.
 For the distribution of outgoing links no analytical results are available.
Our numerical results in the inset to Fig.\ 1 can be well approximated
with the linear dependence  $\tau _{out} \approx 2 + 3\alpha $. Therefore,
the difference between the two exponents $\Delta \equiv \tau _{out}-
\tau _{in} \approx 2\alpha $ persists for any finite $0< \alpha < 1$.
Comparing these results with the available empirical data we  estimate
the parameter $\alpha = 0.22 \pm 0.1$. Here we used the following
values as the experimental estimates for the exponents
$\tau_{in}=2.16 \pm 0.1$, obtained as the
average value from the data in \cite{HTTP} and last reference in \cite{BJA},
and $\tau_{out}=2.62\pm 0.1$ from data in \cite{HTTP}, by fitting only
the straight part of the curves while avoiding the noisy tails and the
curvature at small $q$. It is clear that these values are not definitive.
More careful fits of the data and further measurements are necessary
in order to reduce the experimental error bars.
Hence, the approximative value of the control parameter $\beta $
in the current state of the Web is
$\beta \approx 3$, i.e., in the average  three updated links
come to one added link at each evolution  step.   We now discuss
the properties of the network for this particular value of the control
parameter $\beta$.

\section{Dynamic fractal structures}

In the dynamical systems with large-scale organization the occurrence
of algebraically decaying distributions can be linked to the formation of
certain fractal structures  \cite{fractals} in the system. In the case
of complex evolving networks, however, the existence of such structures
is less clear, due to the spatial randomness of the graph.
To show that the power-law decay of the
distributions of node ranks, shown in Fig.\ 1, is associated with
 certain fractal structures on the network
we examine the {\it temporal pattern} of linking between nodes.
 A part of the pattern is shown in  Fig.\ 2 (top panel).
Clustering in the patterns  indicate frequently active nodes, whereas
voids of various sizes indicate  lesser activity at corresponding  nodes.
Visually heterogeneous  picture suggests that both pattern of origins
of links and pattern of targets have fractal structure.
The fractality of these patterns can be quantified, for instance, by
computing the distribution of time intervals between
two successive linking at a given node (return time).
The distribution of time
intervals $\Delta t$ between two successive linking {\it to} a selected
node in the network $P(\Delta t)$ measured  for
the network of $10,000$ nodes and the parameter $\beta =3$ is given
in Fig.\ 2 (lower panel).  The algebraic decay of the distribution
is a signature of the underlying fractal structure of the pattern.
The distribution of time intervals between two successive
links {\it from} a selected node also shows a power-law tail for large
$\Delta t$. At  small intervals  $\Delta t$ the pattern is nearly random.
The two distributions coincide, indicating mutual correlations in
the patterns, for time intervals $\Delta t > 100$ (that
are accessible for large evolution times $t$).
To emphasize the relation  to Fig.\ 1
we  recall that the number of
links of either kind at a node  accumulates with time.
Hence, the regions of
{\it large time intervals} between successive linking at a
node, $\Delta t$   in Fig.\ 2, correspond to {\it small number
of links}  at that node,  $q$  in Fig.\ 1, and vice versa.

\section{Dynamic critical states}

The scaling behavior of the rank distributions can be examined in terms
of the potential dynamic critical states by measuring the
statistics of triggered avalanches \cite{SPA} on the network.
We consider an analog of the avalanche size in the
networks: the size of a cluster of nodes  which are {\it physically
accessible along directed links}  starting from a random  node in
the network.
A  network is grown using the rules in Eqs.\
(\ref{outlinking})-(\ref{inlinking}). We then select  a random node and
 make a list of nodes  which are connected by outgoing links from that
node. In the next step we make a new list following outgoing links
from the nodes on the previous list, keeping only the
nodes which are different in the two lists, and so on.
The process ends up when no new nodes can be reached---the list is empty.
The number of different lists before an empty list occurs  can be
 recognized as the depth of the connected component.
Technically,  we preserve  ranks $q_{in}$ and $q_{out}$
for each node from the growth phase and, in the case of small networks,
the exact physical links between the nodes. In large networks only
 ranks are preserved and links are searched using the rule in
Eq.\ (\ref{outlinking}). For a large ensemble of networks
the results are expected to be statistically the same.
In this way the size of a cluster of linked nodes represents a
response of the system on the random excitation. The process is reminiscent
 of the invasion percolation on a random graph, where a new edge $j$,
which already does not belong to the graph, is invaded along an outgoing
hyperlink $i\to j$ from the node $i$ if the node rank $q_{out}(i)$ exceeds
unity. The probability to add the edge $j$ to the graph is given by
the preference rule in Eq.\ (\ref{outlinking}). In a more familiar case
of the invasion percolation in physical wetting on the lattice
\cite{invasion_percol}
that probability is given by the least resistance rule. To our knowledge
the problem of invasion percolation on random graphs has not been studied
so far \cite{comment}.

The differential distribution of size of connected clusters of nodes
is  given  in Fig.\ 3 for different values of the parameter $\beta $.
The distribution
of depth of the same clusters is given in the inset to Fig.\ 3.
In the case $\beta =3$ we find that small clusters follow a power-law
distribution with the exponent $\tau _s= 2.79 \pm 0.12$, comparable with
the measured \cite{HTTP} value  $\tau_s=2.52$.
(Note that in the case of Internet the corresponding exponent is estimated
 \cite{internet} as $\tau_s^{I} = 1.90$.)
In Fig.\ 3 the occurrence of the  peak in  the distributions  at large
clusters $S_0 \approx 10^2$ indicates that a large number of the avalanches
have approximately same size.
This is a signature of the  existence of a large subset of strongly connected
nodes (a giant component): Once a `crawl' enters  the subset of strongly
connected nodes, it explores it entirely.  By taking a larger
number of nodes  $N$ the position of the
peak moves towards larger values approximately as $S_0\sim N/7$.

Before discussing  the distributions in Fig.\ 3,  we demonstrate
that the network with the hierarchical structure of links is characterized
by a fractal noise.
In the search of a connected cluster described above we examine detailed
 variation of the number of nodes added to the cluster
at each  step of investigation. We obtain the noisy signal shown Fig.\ 4.
Properties of the signal vary with the control parameter of the dynamics
$\beta $. Considering the number of steps as a total elapsed time of
investigation, the Fourier spectrum of the signal is shown in the top panel
in Fig.\ 4. It  shows a region of correlated behavior with power-law decay
 between the upper cut off at high frequencies
and lower cut off (due to finite size of the network). In addition,
the spectrum in the case of supercritical network exhibits a peak
at a characteristic frequency $f_0 $, which is absent
in the case of the critical parameter $\beta =\beta _C(N)$ (cf. Fig.\ 4).

\section{Discussion and conclusions}

In the theory of critical
states \cite{SPA,TD}, appearance of the peak in the distribution of
avalanches, such as the distributions shown in Fig.\ 3, indicates that
the system is supercritical.
In the  critical state the absence of any characteristic scale
is manifested in the purely algebraic decay of the distribution of
cluster size until  a cut-off, which depends on the size of the network $N$.
On the other hand, the form of distribution  in Fig.\ 3 suggests
that a critical point exists that can be reached by varying a
relevant parameter of the dynamics \cite{TD}. In fact, by decreasing the
parameter $\beta $  we see in Fig.\ 3 that
the slope $\tau _s$ decreases  and the peak eventually disappears at
a critical value \cite{comment} of the parameter $\beta = \beta _c
\approx 0.081$. Beyond the critical value of
the control parameter the scaling behavior of the size and depth of
connected components is entirely lost.  In the critical state the
scaling exponents  are (cf. inset to Fig.\ 1):
$\tau _{in} \approx 2.925$, $\tau _s =\tau _d = 1$.
The distribution  of outgoing links shows very sharp decay, the exponent
$\tau _{out}$ is difficult to measure.
It should be stressed that  the absence of the peak in the critical state
suggests that  no giant component can be formed. Rather, the network
consists  here of many small
groups of well interlinked nodes and percolating directed links
 between these groups.

It should be stressed that the special limit of our model
 $\beta = 0$  (i.e., $\alpha =1$),
corresponding to the original model of preferential attraction of
incoming links proposed by Barabasi, Albert and Jeong \cite{BJA},
belongs to an entirely  different  class of scaling behavior as  regards
the distribution of size and depth of connected clusters: these
distributions are flat between lower and upper cutoffs (see Fig.\ 3).
 The only distribution which enjoys a power law in this limit
is the distribution of incoming links, which has the exact \cite{DMS}
exponent $\tau_{in} =3$ (see also inset to Fig.\ 1).
In this case ($\beta = 0$) updates at  already existing nodes are no
longer possible. This feature---freezing of the outgoing links---as well as
quantitative disagreement in the exponent $\gamma$ for distribution of
incoming links, makes the model of Ref.\ \cite{BJA}
inappropriate as a model of the world-wide Web dynamics.
 A realistic network with frozen outgoing links is
the network of scientific citations \cite{sci}, where  physical
links, corresponding to the cited references in already published papers,
remain fixed in time.

In conclusion, we have demonstrated that certain salient features of the
dynamics of the world-wide Web require more careful modeling, compared to
models of a generic complex evolving network with preferential
attraction of links \cite{BJA} and models of random graphs with rewiring
\cite{RandomGraph}. The dynamic structure and functioning  of the
world-wide Web is deeply rooted in the activity of the agents who are
creating the outgoing links. The hierarchical structure of outgoing links,
which is documented by measurements in the real world-wide Web \cite{HTTP},
is related to the bias activity of the agents in the rule
(\ref{outlinking}) of our model.
A random selection of the active agent (see curve (c) in Fig.\ 1) fails
to describe the distribution of outgoing links in the real Web.
Temporal variation  of the outgoing links {\it inside the network}, i.e.,
updates of links,  has twofold consequences on the global structure
of the network: (i) When the updates of the outgoing links are allowed
(i.e., when $\beta $ strictly larger than zero), the structure of
{\it incoming links} qualitatively changes, the exponent
$\tau _{in}(\beta ) < 3$. The distribution of incoming links in our
model coincides with the analytical  results of Dorogovtsev {\it et al.}
\cite{DMS} in the scaling region. In addition,
the distribution of size and depth of connected clusters becomes
hierarchical when $\beta > 0$, in agreement with results  found in the
real Web, which is not the case when no updates are allowed ($\beta =0$).
(ii) Varying the frequency of updates $\beta $ implies changes in
the  Web structure, both in the outgoing and incoming links. This has
immediate impact on the accessibility of nodes.
  The correlations between the outgoing and incoming links
suggests that the {\it local} structure of the network is qualitatively
different compared to the case without updates.
Here we did not study in detail the probability that two nodes are
linked (clustering coefficient).  Rather we studied the  physical properties
of the network which make the background of the observed behavior: fractal
temporal patterns and number of nodes that join the cluster at one time unit.

 By comparison of the simulated results and
the data obtained in the real world-wide Web we obtained a systematic
agreement both for the distributions of outgoing and incoming links and
for the size of connected components when a single control
parameter $\beta $ is fixed to $\beta =3$ within error bars of the data.
To our knowledge no previous model claimed to describe the
world-wide Web achieved such degree of consistency.
This makes us believe that the present simplified model takes into account
some  basic features of the dynamics of Web.  It would be interesting to
estimate $\beta $ by directly measuring the average number of updated links
in the world-wide Web relative to the number of added links originating from
each added node.

The structure of links shown by the distributions in Figs.\ 1 and 2
is closely related to the evolution of the number of links {\it at a given
node} and to other  local properties of the network.
A detailed study  of these  properties within the present model
requires additional work that  remains to be done in the future.
Here we expect, based on the analytical results for the
incoming links of Ref.\ \cite{DMS},  that in our model
the average number of links $<q_{in}(i,t)>$ at a node $i$  will decay in
time $t\gg i$ as $<q(i,t)>\sim (i/t)^{-\gamma _{in}}$, where
the exponent $\gamma _{in}$ is given by the exact scaling relation
\cite{DMS} $\gamma _{in} =1/(\tau _{in}-1)$, leading to
$\gamma _{in} =1/(1+\alpha )\approx 0.86$.
Assuming that the density of the outgoing links at a given node also
exhibits scaling behavior, as Figs.\ 1-2 suggest, we can predict
a slower decay for the average number of outgoing links at a given node.
The expected exponent is  $\gamma _{out} \approx 1/(1+3\alpha )\approx 0.6$.

 The simulation results that we reported here suggest that the emergent
structure of links in the world-wide Web is strongly related to the updating
policy of the agents: who updates and how often.
It remains to understand  the potentially more intricate reverse effect:
how the amount of information currently stored in the Web influences
the conduct of the agents, with  implications on self-tuning of
the control parameter $\beta$.  In our  model the structure of the network at
$\beta >\beta _c(N) $ appears to be supercritical, where $\beta _c(N) <0.1$
in a large network. Therefore, it is plausible to expect that the
evolutionary selected values of $\beta $ will be much smaller than the
currently estimated value $\beta \approx 3$,
if the scenario of self-tuning of the control parameter is active in
the real Web. In this respect the current state of the real Web can be
regarded as a transient rather than a stationary state.

Finally, our results support the conclusion that in the directed graphs
two growth rules are necessary to describe the dynamics of the
outgoing and incoming links, respectively.  In  the case of the
world-wide Web the statistically correlated distributions of outgoing and
incoming links appear as a fundamental feature of the evolution of the Web.
 This conclusion may serve as a starting point for the future modeling of
the real world-wide Web in terms of the master equations.

\acknowledgments
This work  was  supported by the Ministry
of Science and Technology of the Republic of Slovenia. I am grateful to
Deepak Dhar, Vyatcheslav Priezzhev and \'Alvaro Corral for their response
and for the constructive criticism on this work.

\narrowtext

\begin{figure}
\epsfxsize=82mm\epsffile[42 68 507 526]{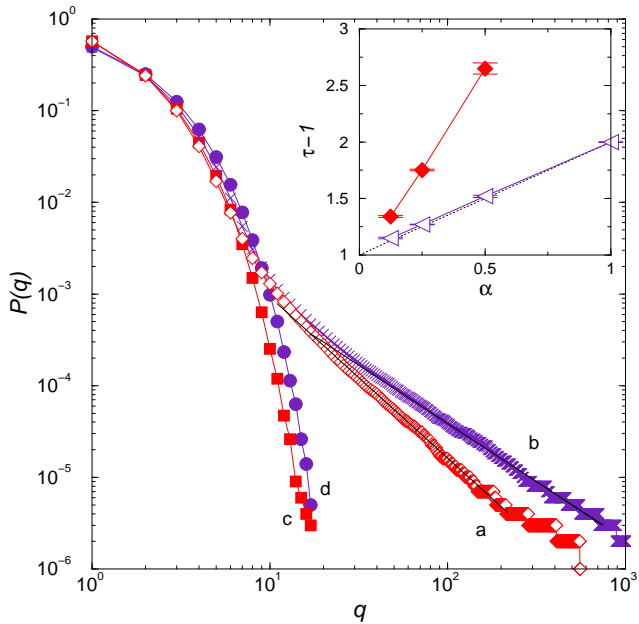}
\vskip 8mm
\caption{\label{fig1}
Cumulative distributions of outgoing links  (a)
and incoming links (b) for the network of $N=1,000,000$ nodes
and average ratio  $\beta =3$ of updated respective to added links
per time unit.   For comparison we have included the corresponding
distributions (c) and (d) in the case of fully random directed graph.
Fitted slopes of the straight sections of the curves (a) and (b)
are compatible with the scaling exponents in Eq.\ (3) as
$\tau_{out}=2.75$ and  $\tau_{in}=2.25$, respectively.
Inset: Variation of the scaling
exponents  $\tau_{out}-1$ (diamonds) and $\tau_{in}-1$ (triangles)
with the control parameter $\alpha \equiv 1/(\beta +1)$ in
the physical range $(0,1)$. Dotted line: exact solution for the case of
incoming links from Ref.\ [11]. }
\end{figure}

\begin{figure}
\epsfxsize=82mm\epsffile[43 67 513 636]{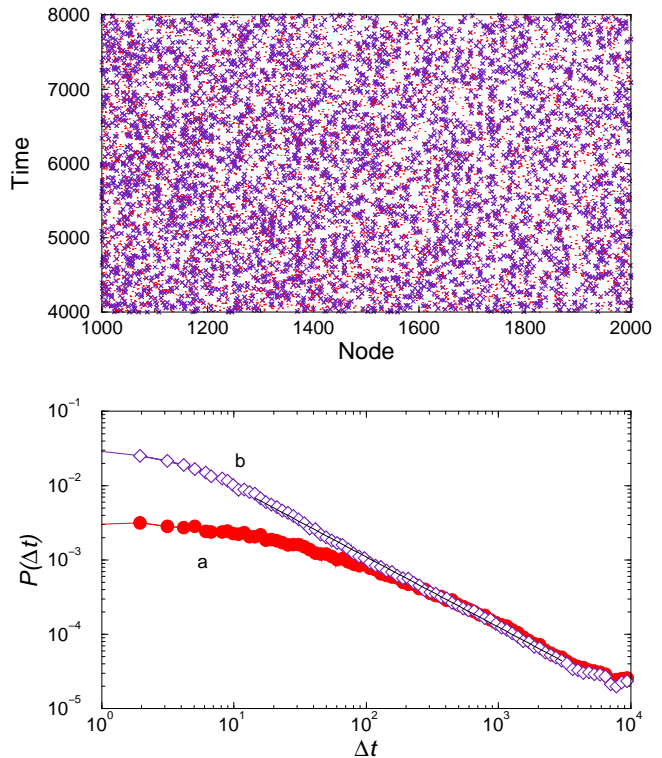}
\vskip 8mm
\caption{\label{fig2}
Top panel: Part of the temporal pattern of
linking in the growth phase of network with $N=10,000$ nodes. One link
per time step is considered. Dots represent nodes from which the link
originates, whereas crosses are target nodes. Lower panel:
Distribution of time intervals
$\Delta t$ between two successive linking  (a) {\it from} a given node,
and  (b) {\it to} a given node. The two patterns become correlated for
the time intervals $\Delta t$ in the range $100 < \Delta t < 3,000$,
corresponding to the scaling region of the rank distributions
$2000 >q>200$ in Fig.\ 1.
 Distributions are averaged over 100 samples and logarithmically binned.
Both distributions are normalized to the total number of time steps. Note
that the events with $\Delta t =0$ that correspond to the outgoing links
from new added nodes, contributing to the distribution (a), are not shown.}
\end{figure}
\vskip 2 true cm
\begin{figure}
\epsfxsize=82mm\epsffile[42 72 565 487]{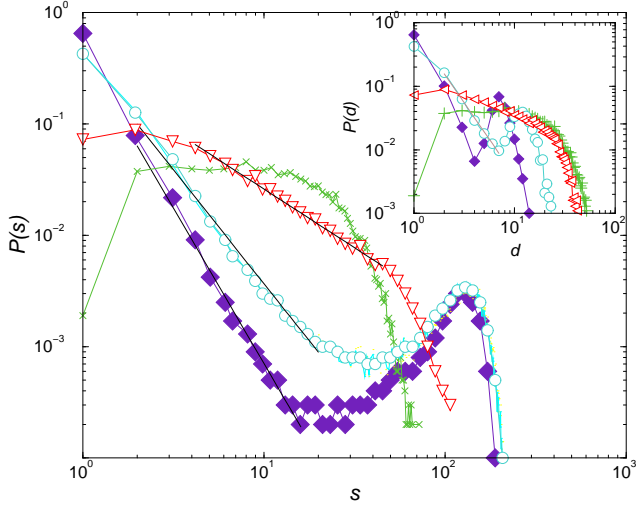}
\vskip 8mm
\caption{\label{fig3}
Differential distribution of size of connected
clusters in the network grown with the rules of Eqs.\
(\ref{outlinking})-(\ref{inlinking}) for different values of
the control parameter $\beta =3$ (diamonds), $1$ (circles), $0.081$
(triangles), and zero (crosses). We employ $1,000$ avalanches in each
of $100$ samples of the network with $N=1,000$ nodes.
Inset: Corresponding distributions of depth of the connected clusters.}
\end{figure}

\begin{figure}
\epsfxsize=82mm\epsffile[67 69 570 634]{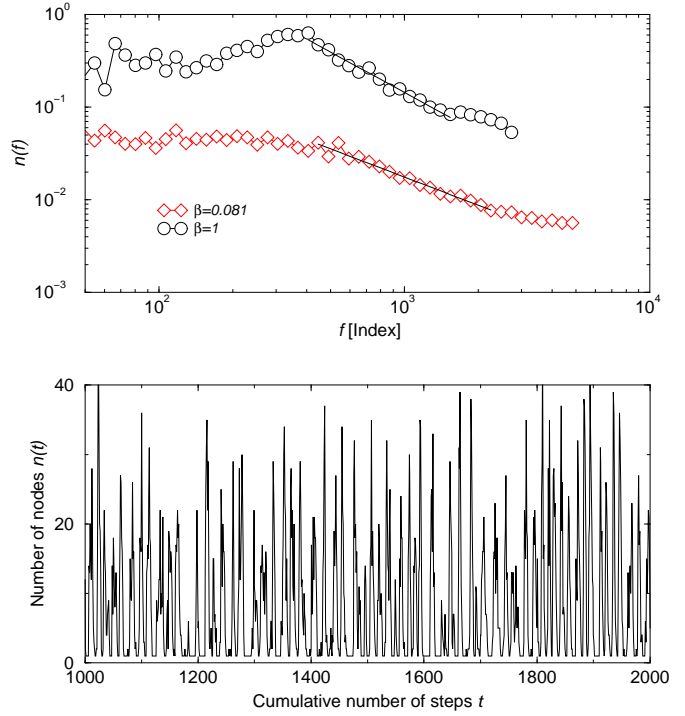}
\vskip 8mm
\caption{\label{fig4}(Lower panel) Number of nodes $n(t)$ added to a
connected cluster at
one investigation step vs cumulative number of steps  $t$. Size of one
connected cluster is represented by the surface enclosed between two
consecutive drops of the signal $n(t)$ to the base line $n(t) =1$.
Parameters are: $N=1,000$ and $\beta = 1$. (Top panel) Fourier spectrum of
the same signal. Also shown is the spectrum of the
corresponding signal at the critical value of the parameter $\beta _c(N)=
0.081$. Data are logarithmically binned.
Straight lines are power-law fits with the slopes $\phi = 1.45 \pm 0.06$
(for $\beta =1$) and $\phi = 1.01 \pm 0.06$ (for $\beta =0.081$).}
\end{figure}

\end{multicols}

\begin{references}
\bibitem{comment}Our results suggest that $0< \beta _c(N) <0.1$ varying with
$N$, but it is strictly larger than zero for any finite $N$ accessible in
the simulations. An estimate of $\beta _c$  by employing different
system sizes and a finite-size scaling analysis remains out of the scope of
this paper.
\end{references}
\end{document}